\def\rfr#1{Equation~\ref{#1}}

\def\eqi{\begin{equation}}
\def\eqf{\end{equation}}
\def\eqia{\begin{eqnarray}}
\def\eqfa{\end{eqnarray}}

\def\lb#1{\label{#1}}

\def\ton#1{\left(#1\right)}

\documentclass[iop,revtx4]{emulateapj}
\usepackage{float}
\usepackage{amsmath,textgreek,wasysym}
\usepackage{amscd,lineno}
\usepackage{amssymb}
\usepackage{graphicx,epsfig}
\usepackage{txfonts}
\usepackage{ulem}
\RequirePackage{color}

\allowdisplaybreaks[1]

\begin{document}

\title{On the post-Keplerian corrections to the orbital periods of a two-body system and their 
application to the Galactic Center}

\author{Lorenzo Iorio$^{1}$, Fupeng Zhang$^{2}$}
\affil{$^{1}$Ministero dell'Istruzione, dell'Universit\`{a} e della Ricerca
(M.I.U.R.)-Istruzione
\\ Permanent address for correspondence: Viale Unit\`{a} di Italia 68, 70125, Bari (BA),
Italy; lorenzo.iorio@libero.it}
\affil{$^{2}$School of Physics and Astronomy, Sun Yat-Sen University, Guangzhou 510275, China;
zhangfp7@mail.sysu.edu.cn}

\begin{abstract}
Detailed numerical analyses  of the orbital motion of a test particle around a spinning
primary are performed. They aim to investigate the possibility of using the post-Keplerian
(pK) corrections to the orbiter's periods (draconitic, anomalistic and sidereal) as a further
opportunity to perform new tests of post-Newtonian (pN) gravity. As a specific scenario,
the S-stars orbiting the Massive Black Hole (MBH) supposedly lurking in Sgr A$^\ast$ at the
center of the Galaxy is adopted. We, first, study the effects of the pK Schwarzchild,
Lense-Thirring  and quadrupole moment accelerations experienced by a target star for
various possible initial orbital configurations. It turns out that the results of the
numerical simulations are consistent with the analytical ones in the small eccentricity
approximation for which almost all the latter ones were derived. For highly elliptical orbits,
the size of all the three pK corrections considered  turn out to increase remarkably. The
periods of the observed S2 and S0-102 stars as functions of the MBH's spin axis orientation
are considered as well. The pK accelerations considered lead to corrections of the
orbital periods of the  order of $1-100~\textrm{d}$  (Schwarzschild), $0.1-10~\textrm{h}$
(Lense-Thirring) and $1-10^3~\textrm{s}$  (quadrupole) for a target star with
$a=300-800~\textrm{AU}$ and $e \approx 0.8$, which could be possibly  measurable by the
future facilities.
\end{abstract}

%
\section{Introduction}
The periodic evolutions of the orbits of a binary system can be described differently by some characteristic orbital
time spans that pertain the crossing of different locations in the sky: the draconitic, anomalistic and
sidereal orbital periods, which are measured according to two consecutive
crossings of the lines of the nodes, periapsis, and a given reference direction in the sky plane,
respectively~\citep{2005som..book.....C}. In the purely Keplerian case, all of them
coincide, while the degeneration is removed when additional accelerations with respect to the Newtonian monopole are
present in the equations of motion. Thus, measuring them would, in principle, provide
a further way to test post-Newtonian (pN) gravity in several independent scenarios. Recently, the post-Keplerian (pK)
corrections to the two-body sidereal, draconitic and anomalistic orbital periods induced by some pK Newtonian and
pN accelerations were analytically calculated in a perturbative framework~\citep{2016MNRAS.460.2445I}.
These pK corrections of the orbital periods are potentially measurable in some astronomical and astrophysical
environments like Earth's satellites~\citep[e.g.,][]{Kassimenko1966,2001PhLA..292...49M}, Solar and
exoplanetary systems~\citep{2016MNRAS.460.2445I}.


The stellar system made of the so-called ``S-stars'' which orbit the Massive Black Hole (MBH) in the Galactic Center
(GC) provide an unique test-bed for probing general relativity (GR) in the strong gravity regime. Since 1992, the
S-stars in the GC have been detected and traced by the Very Large Telecope
(VLT)~\citep{2009ApJ...692.1075G}, New Technology Telescope (NTT) and Keck telescope~\citep{2008ApJ...689.1044G}. These
observations have provided so far accurate constraints on the mass and distance of the Galactic MBH. In
the near future, the GRAVITY facility \citep{2010SPIE.7734E..0YG} on the Very Large Telescope Interferometer (VLTI),
the Thirty Meter Telescope (TMT) \citep{2015RAA....15.1945S}, and the European Extremely Large Telescope (E-ELT)
\citep{2011PASP..123.1334V}, which are expected to perform extremely accurate measurements of the stars' position and
redshift, can be able to reveal the GR effects hidden in their
tracked motion~\citep[e.g.,][]{2015ApJ...809..127Z,2010ApJ...720.1303A,2010ApJ...711..157A,2011ApJ...734L..19A,
2010PhRvD..81f2002M,2011MNRAS.411..453I, 2011PhRvD..84l4001I,2016ApJ...827..114Y}. In the
meantime, the pK corrections of the aforementioned periods could be measurable if the evolutions of the orbital
elements can be inferred from these observables.

Here, we numerically investigate the pK corrections of S-stars' orbital periods under various initial
conditions and discuss their possible applications for GR tests. The paper is organized as follows. In
Section \ref{metodo}, we briefly describe the adopted pN accelerations and the numerical strategies in obtaining the
different orbital periods. Section \ref{risultati} collects the results obtained for the Schwarzschild-like,
Lense-Thrring and the quadrupole momentum effects. Section~\ref{discussion} provides a discussion of the
results obtained, which are finally summarized in Section~\ref{fine}.
\section{The numerical method used}\lb{metodo}
In our numerical simulations, we integrate the equations of motion of a test star moving around a central black hole
under the action of the Newtonian and pN accelerations. The standard Newtonian monopole is given by
$a_\textrm{N} =-GM/r^2$, where $M$ is the mass of the central black hole, $G$ is the gravitational constant, and $r$ is
the distance of the test particle from $M$. In this work, we also consider the effects of following pK accelerations
\citep{1989racm.book.....S, 1991ercm.book.....B} of both Newtonian and pN origin:

\begin{itemize}
\item The pN Schwarzschild-type gravitoelectric (GE) acceleration due to the
static mass:
\begin{equation}\begin{aligned}
\mathbf{\mathbf{a}}_\textrm{GE}=&-\frac{GM}{r^3 c^2}\left[
\left(v^2-4\frac{GM}{r}\right)\mathbf{r} - 4(\mathbf{r}\cdot \mathbf{v}) \mathbf{v}\right],\\
\label{eq:a1pn}
\end{aligned}\end{equation}
which is the largest 1pN term in the pN formalism. Here, $c$ is the speed of light, $\mathbf{r}$ and
$\mathbf{v}$ are the position and velocity vectors of the star, respectively, while $v=|\mathbf{v}|$ is
its speed.

\item The pN Lense-Thirring gravitomagnetic acceleration due to the slow rotation of the central body:
\begin{equation}\begin{aligned}
\mathbf{a}_\textrm{LT}&=\frac{2GS}{c^2r^3}\left[3(\hat{\mathbf{S}}\cdot\hat{\mathbf{r}})\hat{\mathbf{r}}\times\mathbf{v}
+\mathbf{v}\times\hat{\mathbf{S}}\right],
\label{eq:alt}
\end{aligned}\end{equation}
where $\hat{\mathbf{r}}$ is the versor of the position vector.
In the BH case, the angular momentum is given by $S =\chi M^2 G c^{-1}$, where $\chi$ is the dimensionless spin
parameter, while $\hat{\mathbf{S}}$ is the unit vector of the spin.

\item The Newtonian acceleration due to the quadrupole mass moment $J_2$ of the primary:
\begin{equation}
\begin{aligned}
\mathbf{a}_{J_2} &=
\frac{J_2 GM R_e^2}{2r^4}\left\{\left[
5(\hat{\mathbf{S}}\cdot \hat{\mathbf{r}})^2-1\right]\hat{\mathbf{r}}-
2(\hat{\mathbf{S}}\cdot \hat{\mathbf{r}})\hat{\mathbf{S}}
\right\}.
\label{eq:aj2}
\end{aligned}
\end{equation}
For a rotating BH, an acceleration analogue to Equation \ref{eq:aj2} arises in a general relativistic context in such a
way that $J_2  GM R_e^2 \rightarrow Q = -\chi^2 G^3 M^3 c^{-4} = -S^2 G c^{-1} M^{-1}$
\citep{1970JMP....11.2580G,
1974JMP....15...46H}, and $q = -\chi^2$
is the dimensionless quadrupole moment.
\end{itemize}

The six orbital elements $a$ (orbital semimajor axis), $e$ (eccentricity), $I$ (Inclination to the reference $\{x,~y\}$
plane which coincides with the plane of the sky as in Fig. 1 of \citep{2015ApJ...809..127Z}), $\Omega$ (longitude of the
ascending node), $\omega$ (argument of the periapsis), $f$ (true anomaly) of the target star are calculated according to
the state vector of  the star: $x,y,z$ and $\dot x,\dot y,\dot z$ at any instant. Then, the different periods can be
obtained as follows: (1) The draconitic period is the time interval between $u=0$ and $u=2\pi$; here, $u=\omega+f$ is
the argument of latitude. (2) The anomalistic is the time interval between
$f=0$ and $f=2\pi$. (3) The sidereal period is the time interval between $l=0$ and $l=2\pi$, where
$l=\Omega+\omega+f$ is the true longitude~\footnote{We notice that the results are the same if the intervals are
measured between $u=u_0$ and $u=u_0+2\pi$ for arbitrary $u_0$, similarly for $f$ and $l$.}. We denote $P_{\rm dra}$,
$P_{\rm ano}$ and $P_{\rm sid}$ as the draconitic, anomalistic and sidereal periods, respectively.

Each effect of the Schwarzchild, Lense-Thirring or quadrupole moment can be calculated by
switching the corresponding accelerations showed in Equation~\ref{eq:a1pn}-\ref{eq:aj2} on and off
and, then, inspect the difference of these periods. The perturbed period by these effects are
denoted as $\delta_\textrm{GE} P$, $\delta_\textrm{LT} P$ and $\delta_{Q} P$, respectively. Here, $P$
could be any of $P_{\rm dra}$, $P_{\rm ano}$ or $P_{\rm sid}$.

We adopt the code DORPRI5 based on the explicit fifth (fourth)-order Runge Kutta
method~\citep{1980Dormand19,1993Hairer} to integrate the equations of motion to obtain the orbital path of the star. The
relative integration error is $\le10^{-12}$ for all the simulations performed in this work, which is sufficient for the
convergence of the numerical results in this study.

Note that, in some cases, the orbit of the target star is initially bound in Newtonian
gravity but turns out to be unbound while the pK corrections are considered,
especially for those stars with high initial
orbital eccentricities. We avoid these unbound orbits as they do not have periodic
patterns. In the numerical simulations, we remove the stars that initially have
positive specific orbital energy, i.e, $E_{\rm orb}>0$. Here $E_{\rm orb}$ is given by~\citep{1995PhRvD..52..821K}
\begin{equation}
E_{\rm
orb}=\frac{1}{2}v^2-\frac{GM}{r}+\frac{3v^4}{8c^2} +
\frac{3v^2}{2c^2}\frac{GM}{r}+\frac{G^2M^2}{2c^2r^2}.
\end{equation}
\section{Results}\lb{risultati}
First, for each of the pK effects considered, we calculate the time series of the element $u(t),~f(t),~l(t)$, with and
without the pK accelerations by adopting the same initial conditions for the target star in both the integrations. Then,
for each run, we estimate the periods $P$ according to the resulting numerically integrated time series of
$u(t),~f(t),~l(t)$. Finally, we get the difference $\delta P$ of such periods between each couple of runs, which shows
the effects of the corresponding pK acceleration.

In the following, we will assume that the mass of the central MBH is $M_\bullet = 4\times10^6 M_\odot$, which
approximates the one found in the Galactic center~\citep[e.g.,][]{2009ApJ...692.1075G,2008ApJ...689.1044G}.
As far as the orbital geometries of the test particle adopted are concerned, we will consider both fictitious orbital
configurations and those of the so-far discovered S2 ($a_0 = 984~\textrm{AU},~e_0 = 0.88,~I_0 = 135^\circ,~\Omega_0 =
225^\circ,~\omega_0 = 243^\circ$) and S0-102 ($a_0 = 848~\textrm{AU},~e_0 = 0.68,~ I_0 = 151^\circ,~\Omega_0 =
175^\circ,~\omega_0 = 5^\circ$). Since the gravitoelectric anomalistic correction (cfr. Equation (72) of
\citet{2016MNRAS.460.2445I} ) and all the three corrections due to the quadrupole mass moment (cfr. Equations (65) to
(67) of \citet{2016MNRAS.460.2445I}) depend on $f_0$, we adopted in our simulations the values $f_0 = 157^\circ$ (S2)
and $f_0 = 225^\circ$ (S0-102), which correspond to the initial orbital position of S2 and S0-102 in the year 2020,
respectively.

The spin axis of the MBH are parameterized by the angle $i$ and $\epsilon$ as
\begin{align}
{\hat{S}}_x & = \sin i\cos\epsilon, \\ \nonumber \\
{\hat{S}}_y & = \sin i\sin\epsilon, \\ \nonumber \\
{\hat{S}}_z & = \cos i,
\end{align}
with $0\leq i\leq \pi,~0\leq\epsilon\leq 2\pi$ by considering the angles $i,~\epsilon$ as free parameters.
Here, $i$ is the angle of the spin with respect to the line of sight, and $\epsilon$ is the angle
between the projected direction of the spin in the plane of the sky and the reference $x-$axis.

Currently, the spin of the MBH in the GC remains largely unconstrained. The radio
observations of Sgr A* provide
a weak constraint of the spin parameters of the MBH~\citep{2009ApJ...697...45B}, with $1\sigma$ estimated values given
by $\chi<0.4$, $i={50^\circ}^{+10^\circ}_{-10^\circ}$ and
$\epsilon={-20^\circ}^{+31^\circ}_{-16^\circ} $. It is also speculated that the spin axis of the MBH
could possibly coincide with the orientation axis of the young stellar disk in the GC, if the latter is the remnant of
a previously existing accretion disc~\citep{2016ApJ...827..114Y}. In this case, the
possible spin orientations are $i=130^\circ$, $\epsilon=6^\circ$ or $i=50^\circ$, $\epsilon=186^\circ$. In this work,
we consider both general orientations of the spin axis and the three aforementioned values.

\subsection{The 1pN gravitoelectric Schwarzchild-like effect due to a non-rotating primary}\label{subsec:GEsec}
\begin{figure*}
\centering
\includegraphics[scale=0.55]{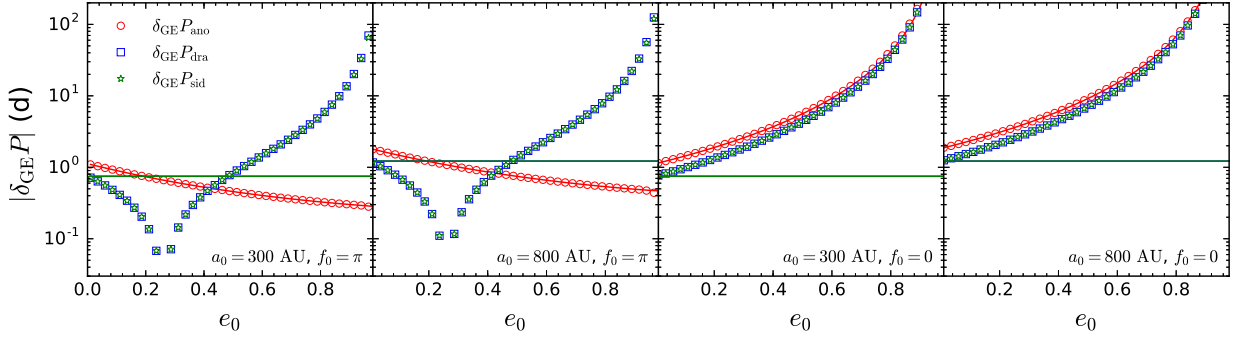}
\caption{Perturbed orbital periods, in d, due to the 1pN static Schwarzchild-type
gravitoelectric field of the MBH. The red open circle, blue square and green star symbols  show the simulation's results
for the perturbed
anomalistic $\delta_\textrm{GE} P_{\rm ano}$, draconitic $\delta_\textrm{GE} P_{\rm dra}$, and sidereal
$\delta_\textrm{GE} P_{\rm sid}$ periods, respectively. Different panels show when $a_0$ and $f_0$ take different values
(See the legend in the bottom right of each panel). The
red, blue and green solid lines show the  values theoretically predicted by Equations~\ref{eq:ge}. The blue and green
lines, which hold to zero order in the eccentricity, overlap with each other as they have the same value.}
\label{fig:dpsw}
\end{figure*}
The first order theoretical predictions of the 1pN corrections to the orbital period
are given by (cfr. Equations 69-72 of \citet{2016MNRAS.460.2445I})
\begin{equation}
\begin{aligned}
\delta_{\rm GE}P_{\rm dra}&=\frac{12\pi\sqrt{\mu a}}{c^2}+\mathcal{O}(e^n),~n\ge 1,\\
\delta_{\rm GE}P_{\rm ano}&=\frac{3\pi\sqrt{\mu
a}}{c^2(1-e^2)^2}\mathcal{T}_\textrm{ano}^\textrm{(GE)},\\
\delta_{\rm GE}P_{\rm sid}&=\frac{12\pi\sqrt{\mu a}}{c^2}+\mathcal{O}(e^n),~n\ge 1.
~\label{eq:ge}
\end{aligned}
\end{equation}
Here,
\begin{equation}
\mathcal{T}_\textrm{ano}^\textrm{(GE)} = 6+7e^2+2e^4+2e(7+3e^2)\cos f_0+5e^2\cos
2f_0,
\end{equation} and $\mu=GM$. Note that such an effect increases as a function of the distance of the star to the MBH
($\propto\sqrt{a}$). Only analytical expressions for $e=0$ are derived for $\delta_{\rm GE} P_{\rm dra}$,
$\delta_{\rm GE} P_{\rm sid}$ and $\delta_{\rm GE} P_{\rm ano}$ while $\delta_{\rm GE} P_{\rm ano}$ is obtained
explicitly for arbitrary eccentricities.

To explore corrections to the periods induced by the gravitoelectric, Schwarzschild-type acceleration for target stars
with various initial conditions, we perform the numerical simulations for an exemplified fictitious target star with the
following initial conditions: $a_0=300~\textrm{AU}$ or $800~\textrm{AU}$, $f_0=0$ or $f_0=\pi,~I_0=\pi/4,$ $\Omega_0=0$,
$\omega_0=\pi$, while the eccentricity $e_0$ is allowed to continuously vary from  $0.01$ to $0.95$.
Figure~\ref{fig:dpsw} depicts the results of $\delta_\textrm{GE} P_\textrm{dra}\ton{e_0},~\delta_\textrm{GE}
P_\textrm{ano}\ton{e_0},~\delta_\textrm{GE} P_\textrm{sid}\ton{e_0}$ as functions of $e_0$. It can be noted that the
agreement with the theoretical predictions of \citet{2016MNRAS.460.2445I} is remarkable, especially in the case of the
anomalistic period for which an exact analytical expression with respect to the eccentricity is available. As far as
the draconitic and the sidereal periods are concerned, Equation~\ref{eq:ge} yield identical predictions for both, valid
to zero order in the eccentricity. Our numerically integrated results agree with them in the
limit of small values of $e_0$. As we can see from Figure~\ref{fig:dpsw}, the period corrections can vary
by $1\sim2$ orders of magnitude for different values of $e_0$ if $e_0\lesssim 0.95$.

According to Equation~\ref{eq:ge}, which is also confirmed by numerical simulations, the anomalistic
corrections doest not depend on the orbital orientation determined by $I,~\Omega,~\omega$. For the
draconitic and sidereal corrections $\delta_\textrm{GE} P_\textrm{dra},~\delta_\textrm{GE} P_\textrm{sid}$, however, we
found that they  depend largely on the these parameters. Note that in Figure~\ref{fig:dpsw} the draconitic corrections
are equal to that of the sidereal ones, i.e., $\delta_\textrm{GE} P_\textrm{dra}=\delta_\textrm{GE}
P_\textrm{sid}$;it is a coincidence as we initially set $\Omega_0=0$. We find that $\delta_\textrm{GE}
P_\textrm{dra}\ne \delta_\textrm{GE} P_\textrm{sid}$ if $\Omega_0\ne0$. By performing a large number of numerical
simulations, we find that the pK effects of all the three-types of periods can vary by $\sim 1$ orders of magnitude if
we select arbitrary values of $I_0,~\Omega_0,~\omega_0,~f_0$.

For the currently detected S-star S2 and S0-102, we can obtain their explicit values of the three-types of period
corrections. For S2, we have
\begin{align}
\delta_\textrm{GE}P_{\rm ano} \lb{ano1} & = 1.4~\textrm{d}, \\ \nonumber \\
\delta_\textrm{GE}P_{\rm dra} & = 0.51~\textrm{d}, \\ \nonumber \\
\delta_\textrm{GE}P_{\rm sid} & = 0.78~\textrm{d}.
\end{align}
while for S0-102 our results are
\begin{align}
\delta_\textrm{GE}P_{\rm ano} \lb{ano2} & = 1.7~\textrm{d}, \\ \nonumber \\
\delta_\textrm{GE}P_{\rm dra} & = 1.6~\textrm{d}, \\ \nonumber \\
\delta_\textrm{GE}P_{\rm sid} & = -2.8~\textrm{d}.
\end{align}
Remarkably, \rfr{ano1} and \rfr{ano2} agree with the corresponding values calculated analytically with
Equation~\ref{eq:ge}.
\subsection{The 1pN gravitomagnetic Lense-Thirring effect due to the angular momentum of the
primary}\label{subsec:LTsec}
\begin{figure*}
\centering
\includegraphics[scale=0.55]{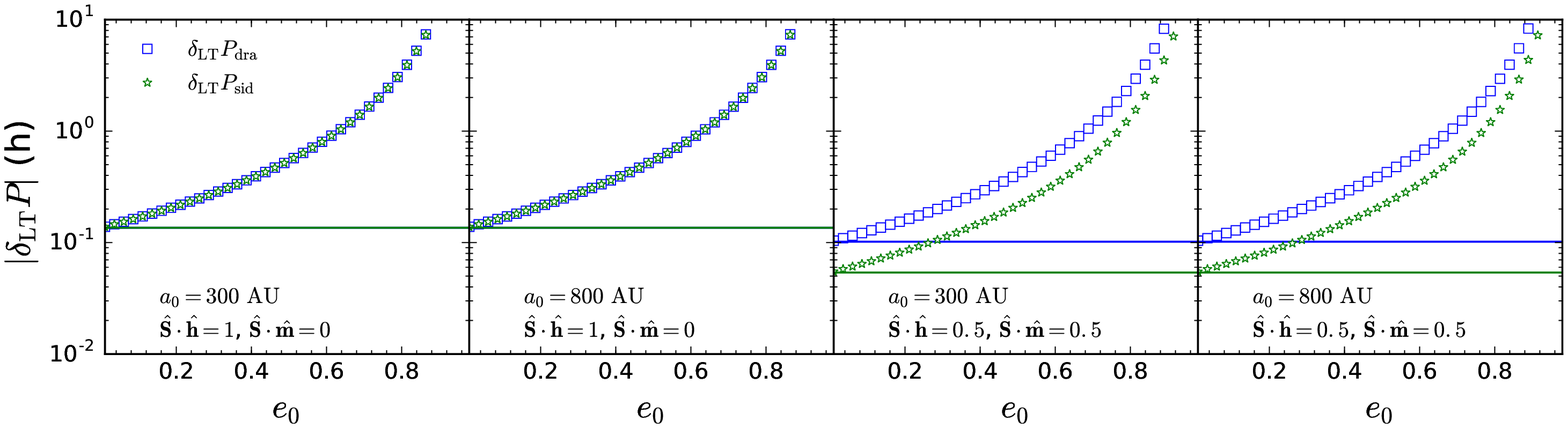}
\caption{
Perturbed orbital periods, in h, due to the 1 pN stationary Lense-Thirring
gravitomagnetic field of the MBH as functions of the orbital eccentricity $e_0$. The blue open square and
the green star symbols show the simulation's results for the perturbed draconitic $\delta_{\rm LT} P_{\rm dra}$ and
sidereal  $\delta_{\rm LT} P_{\rm sid}$ periods,
respectively. Different panels show the simulation's results when $a_0$ and the spin--orbit orientation take
different values (See the legend in the bottom of each panel). The continuous blue and green lines show
the values theoretically predicted by Equation~\ref{eq:lt}.}
\label{fig:dplt}
\end{figure*}
\begin{figure*}
\center
\includegraphics[scale=0.6]{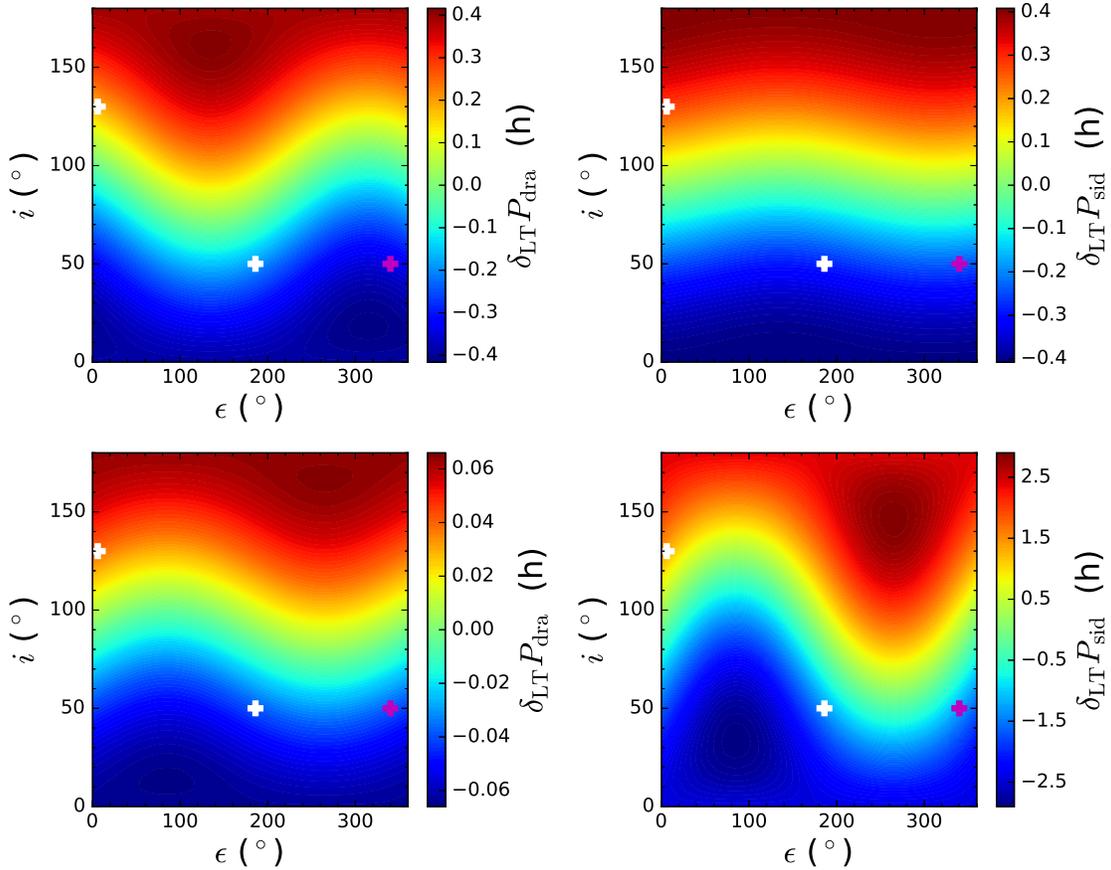}
\caption{Upper row: draconitic and sidereal corrections for the S2 star due to the Lense-Thirring effects as functions
of the $i,~\epsilon$ angles characterizing the spatial orientation of the spin axis of the MBH in the GC.
The write and magenta cross show the spin orientation suggested by that of the young stellar disc in the
GC~\citep{2016ApJ...827..114Y} and those from~\citep{2009ApJ...697...45B}, respectively. Lower row: same as in the upper
row for the
S0-102 star.}
\label{S2S0102_LT_periods}
\end{figure*}
The theoretical predictions of the 1pN gravitomagnetic corrections to the orbital period
are given by (cfr. Equations 74-78 of \citet{2016MNRAS.460.2445I})
\begin{equation}
\begin{aligned}
\delta_{\rm LT}P_{\rm dra}&=\frac{4\pi S }{c^2
M}\left[2\left(\hat{\mathbf{S}}\cdot\hat{\mathbf{h}}\right)+\left(\hat{\mathbf{S}}\cdot\hat{\mathbf{m}}\right)\cot
I\right]+\mathcal{O}(e^n),~n\ge 1,\\
\delta_{\rm LT}P_{\rm ano}&=0,\\
\delta_{\rm LT}P_{\rm sid}&=\frac{4\pi S}{c^2
M}\left[2\left(\hat{\mathbf{S}}\cdot\hat{\mathbf{h}}\right)-\left(\hat{\mathbf{S}}\cdot\hat{\mathbf{m}}\right)\tan
I\right]+\mathcal{O}(e^n),~n\ge 1.
\label{eq:lt}
\end{aligned}
\end{equation}
Here, $\hat{\mathbf{m}}$ is the unit vector directed transversely to the line of the nodes in the orbital
plane, and $\hat{\mathbf{h}}$ is the unit vector of the orbital angular momentum per unit mass of the test
particle \citep[for more details, see][]{2016MNRAS.460.2445I}.
Note that the above equations are independent of $a_0$.

The effects in case of various initial conditions of the stellar orbit and the MBH spin can be explored numerically.
Let us assume that, initially, $a_0=300~\textrm{AU}$ or $a_0=800~\textrm{AU}$,
$I_0=\pi/4$, $\Omega_0=0$, and  $\omega_0=f_0=\pi$, while $e_0$ varies continuously from $0.01$ to $0.95$.
As far as the magnitude of the MBH's angular momentum is concerned, we adopt $\chi=1$. To avoid
possible misunderstandings, in this section we consider the rotation of the primary to the lowest order in its angular
momentum $\mathbf{S}$, i.e. only frame-dragging or the Lense-Thirring effect; we do not consider the quadrupole of the
MBH which, according to the no-hair theorems \citep{1970JMP....11.2580G, 1974JMP....15...46H}, is of higher order in
the angular momentum, being proportional to its square.
The spatial orientations of the MBH's spin and of the stellar orbital plane are as such:
\begin{itemize}
\item The spin direction is perpendicular to the orbital plane of the star, i.e.,
$\hat{\mathbf{S}}\cdot\hat{\mathbf{h}}=1$,
$\hat{\mathbf{S}}\cdot\hat{\mathbf{m}}=0$;
\item The spin direction is tilted to the orbital plane of the star, i.e.,
$\hat{\mathbf{S}}\cdot\hat{\mathbf{h}}=0.5$,
$\hat{\mathbf{S}}\cdot\hat{\mathbf{m}}=0.5$.
\end{itemize}

The results are showed in Figure~\ref{fig:dplt} in which $\delta_\textrm{LT}
P_\textrm{dra}\ton{e_0}$, $\delta_\textrm{LT} P_\textrm{sid}\ton{e_0}$ are
plotted as functions of $e_0$ for given values of some of the other orbital parameters. In the small eccentricity
regime, for which a direct comparison with the analytical results provided by Equation~\ref{eq:lt} is meaningful,
the agreement is remarkable for both the spin-orbit configurations considered (left and right columns of
Figure~\ref{fig:dplt}). For the anomalistic period, the resulting corrections are not showed as they
are consistent with zero, in agreement with Equation~\ref{eq:lt}. Note that the three pK effects can vary by $1\sim2$
orders of magnitude if we select arbitrary values of $e_0$ ($e_0\lesssim 0.95$), $I_0,~\Omega_0,~\omega_0$.

Equation~\ref{eq:lt} suggests that the gravitomagnetic corrections of the period depend strongly on the
spin orientation. To explore such dependence we
perform numerical simulations for the currently detected stars S2 and S0-102. Figure \ref{S2S0102_LT_periods} depicts
our findings for S2 and S0-102 as functions of the spatial orientation of the spin axis of the Galactic MBH.
It can be noted that, {the effects have both two local maximums in the $i-\epsilon$ space}. For S2,
$\left|\delta_\textrm{LT} P_\textrm{dra},~\delta_\textrm{LT} P_\textrm{sid}\right|\lesssim
0.4~\textrm{h}$, while for S0-102 we have $\left|\delta_\textrm{LT} P_\textrm{dra}\right|\lesssim
0.06~\textrm{h},~\left|\delta_\textrm{LT} P_\textrm{sid}\right|\lesssim 3~\textrm{h}$. If the spin axis is set to
values given by~\citep{2009ApJ...697...45B} and assuming $\chi=0.4$, we find that $\delta_{\rm LT}P_{\rm dra}=-0.34$ h,
and $\delta_{\rm LT}P_{\rm sid}=-0.23$ h for S2 and $\delta_{\rm LT}P_{\rm dra}=-0.04$ h, and $\delta_{\rm LT}P_{\rm
sid}=-1.2$ h for S0-102. Alternatively, if
$i=130^\circ$, $\epsilon=6^\circ$(or $i=50^\circ$, $\epsilon=186^\circ$) and assuming $\chi=1$, we have $\delta_{\rm
LT}P_{\rm dra}=0.19$ h, and $\delta_{\rm LT}P_{\rm sid}=0.28$ h for S2 and
$\delta_{\rm LT}P_{\rm dra}=0.04$ h, and $\delta_{\rm LT}P_{\rm sid}=1.3$ h for S0-102 (or $\delta_{\rm
LT}P_{\rm dra}=-0.19$ h, and $\delta_{\rm LT}P_{\rm sid}=-0.28$ h for S2 and $\delta_{\rm LT}P_{\rm dra}=-0.04$ h,
and
$\delta_{\rm LT}P_{\rm sid}=-1.3$ h for S0-102).

\subsection{The quadrupole moment effects}
The theoretical quadrupole momentum effects red are given by (cfr.
Equations 62-67 of
\citet{2016MNRAS.460.2445I})
\begin{equation}
\begin{aligned}
\delta_{\rm Q}P_{\rm dra}&=\frac{3\pi
Q\mu^{3/2}}{2\sqrt{a}}\mathcal{T}_{\rm dra}^Q+\mathcal{O}(e^n),~n\ge 1,\\
\delta_{\rm Q}P_{\rm ano}&=\frac{3\pi
Q\mu^{3/2}}{2\sqrt{a}}\mathcal{T}_{\rm ano}^Q+\mathcal{O}(e^n),~n\ge 1,\\
\delta_{\rm Q}P_{\rm sid}&=\frac{3\pi
Q\mu^{3/2}}{2\sqrt{a}}\mathcal{T}_{\rm sid}^Q+\mathcal{O}(e^n),~n\ge 1.\\
~\label{eq:q}
\end{aligned}
\end{equation}
Here,
\begin{equation}
\begin{aligned}
\mathcal{T}_{\rm
dra}^Q&=-4+6\left(\hat{\mathbf{S}}\cdot\hat{\mathbf{l}}\right)^2+6\left(\hat{\mathbf{S}}\cdot\hat{\mathbf{m}}\right)^2\\
+&3\left[\left(\hat{\mathbf{S}}\cdot\hat{\mathbf{l}}\right)^2-\left(\hat{\mathbf{S}}\cdot\hat{\mathbf{m}}\right)^2\right
]\cos 2 u_0\\
+&6\left[\left(\hat{\mathbf{S}}\cdot\hat{\mathbf{l}}\right)\left(\hat{\mathbf{S}}\cdot\hat{\mathbf{m}}\right)\right]
\sin 2u_0\\
-&2\left[\left(\hat{\mathbf{S}}\cdot\hat{\mathbf{l}}\right)\left(\hat{\mathbf{S}}\cdot\hat{\mathbf{m}}\right)\right]\cot
I,\\
\end{aligned}
\end{equation}
\begin{equation}
\begin{aligned}
\mathcal{T}_{\rm
ano}^Q&=-2+3\left(\hat{\mathbf{S}}\cdot\hat{\mathbf{l}}\right)^2+6\left(\hat{\mathbf{S}}\cdot\hat{\mathbf{m}}\right)^2\\
+&3\left[\left(\hat{\mathbf{S}}\cdot\hat{\mathbf{l}}\right)^2-\left(\hat{\mathbf{S}}\cdot\hat{\mathbf{m}}\right)^2\right
]\cos 2 u_0+\\
+&6\left[\left(\hat{\mathbf{S}}\cdot\hat{\mathbf{l}}\right)\left(\hat{\mathbf{S}}\cdot\hat{\mathbf{m}}\right)\right]\sin
2u_0,\\
\end{aligned}
\end{equation}
\begin{equation}
\begin{aligned}
\mathcal{T}_{\rm
sid}^Q&=-4+6\left(\hat{\mathbf{S}}\cdot\hat{\mathbf{l}}\right)^2+6\left(\hat{\mathbf{S}}\cdot\hat{\mathbf{m}}\right)^2\\
+&3\left[\left(\hat{\mathbf{S}}\cdot\hat{\mathbf{l}}\right)^2-\left(\hat{\mathbf{S}}\cdot\hat{\mathbf{m}}\right)^2\right
]\cos 2 u_0\\
+&6\left[\left(\hat{\mathbf{S}}\cdot\hat{\mathbf{l}}\right)\left(\hat{\mathbf{S}}\cdot\hat{\mathbf{m}}\right)\right]\sin
2u_0\\
+&2\left[\left(\hat{\mathbf{S}}\cdot\hat{\mathbf{l}}\right)\left(\hat{\mathbf{S}}\cdot\hat{\mathbf{m}}\right)\right]
\tan \left(\frac{I}{2}\right).\\
\end{aligned}
\end{equation}
Here $\hat{\mathbf{l}}$ is the unit vector directed along the line of the nodes toward the ascending
node~\citep{2016MNRAS.460.2445I}.

The above explicit analytical expression hold only for the case $e=0$. Similarly to
Section~\ref{subsec:GEsec} and~\ref{subsec:LTsec},  we perform several numerical simulations
in order to illustrate the effects for target stars with various initial conditions.
Initial conditions are $a_0=100~\textrm{AU}$ or $a_0=300~\textrm{AU}$,
$I_0=\pi/4$,  $\omega_0=f_0=\pi$, so that $u_0=\omega_0+f_0=0$; $e_0$ is allowed to vary continuously from $0.01$ to
$0.95$. As far as the modeled perturbing accelerations are concerned, only the quadrupole term, proportional to $S^2$
in the BH case due to the no-hair theorems, is included; the Lense-Thirring one, linear in $S$, is switched to zero. As
in Section \ref{subsec:LTsec}, it is assumed $\chi=1$. The MBH's spin and the star's orbit are
oriented as
follows:
\begin{itemize}
\item The spin direction is perpendicular to the orbital plane of the star, i.e.,
$\hat{\mathbf{S}}\cdot\hat{\mathbf{h}}=1$,
$\hat{\mathbf{S}}\cdot\hat{\mathbf{m}}=0$, $\hat{\mathbf{S}}\cdot\hat{\mathbf{l}}=0$;
\item The spin direction is tilted to the orbital plane of the star, i.e.,
$\hat{\mathbf{S}}\cdot\hat{\mathbf{h}}=0.5$, $\hat{\mathbf{S}}\cdot\hat{\mathbf{m}}=0.5$,
$\hat{\mathbf{S}}\cdot\hat{\mathbf{l}}=-\cos \frac{\pi}{4}$.
\end{itemize}

\begin{figure*}
\centering
\includegraphics[scale=0.55]{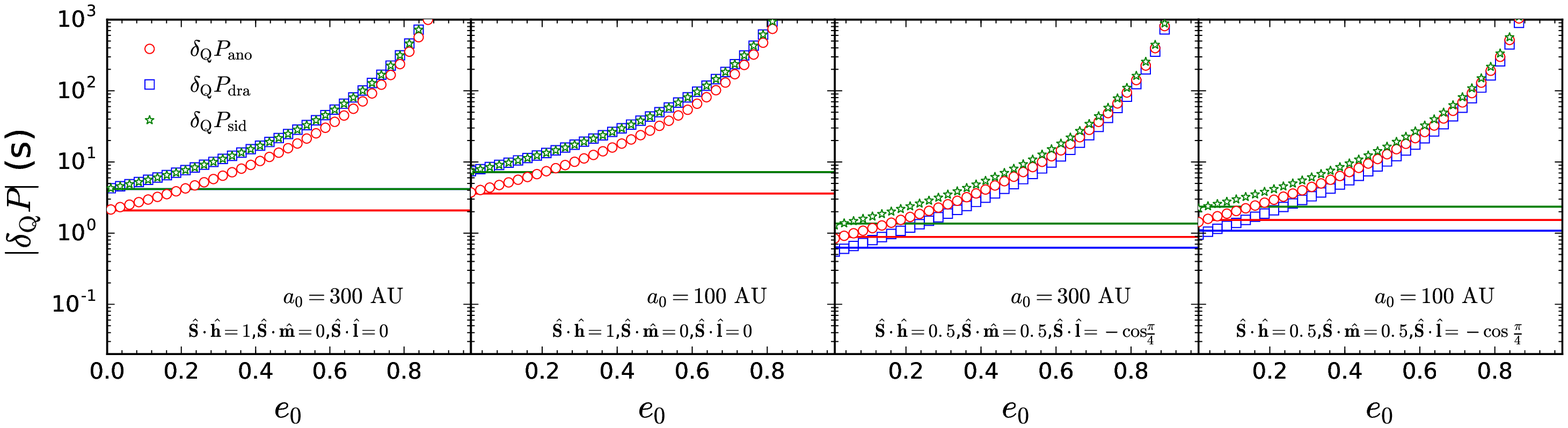}
\caption{Perturbed orbital periods due to the MBH's quadrupole moment (in s) which, according to the no-hair theorems,
is of second order with respect to the angular momentum. The red open circle, blue square and green star symbols show
the simulation's results for the perturbed
anomalistic $\delta_{Q} P_{\rm ano}$, draconitic $\delta_{Q} P_{\rm dra}$, and sidereal
 $\delta_{Q} P_{\rm sid}$ periods, respectively. Different panels show the simulation's results
when $a_0$ and the spin--orbit configuration keep different values (See the legend in
the bottom of each panel). The red, blue and green continuous lines show the theoretical predictions by Equation
62--64 of \citet{2016MNRAS.460.2445I}, valid to zero order in eccentricity.
}
\label{fig:dpqd}
\end{figure*}
The results are showed in Figure~\ref{fig:dpqd}.
A comparison with the analytical predictions of Equations~\ref{eq:q}, accurate to zero order
in eccentricity, with our numerical results for small values of $e_0$ shows a good agreement for both the spin--orbit
configurations considered (left and right columns of Figure \ref{fig:dpqd}). For more eccentric orbits, the draconitic
and the sidereal corrections keep their equality if the MBH's spin is perpendicular to the orbital plane (left column of
Figure \ref{fig:dpqd}), while they are different for the generic spin-orbit configuration adopted in the right column of
Figure \ref{fig:dpqd}. In both cases, the anomalistic correction does not coincide with the other two.

\begin{figure*}
\center
\includegraphics[scale=0.45]{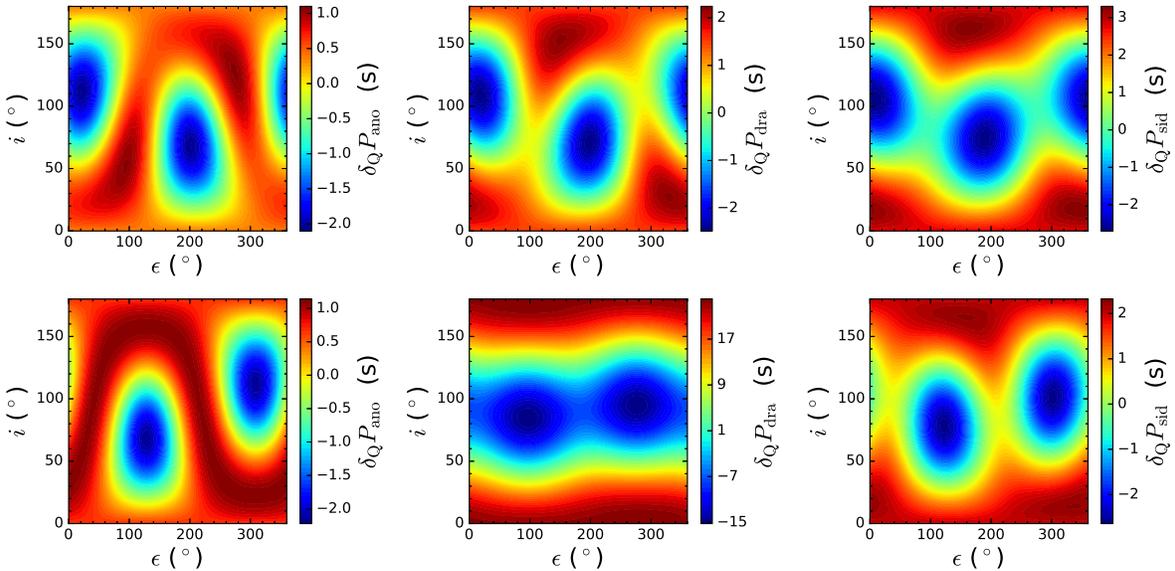}
\caption{
Upper row: anomalistic, draconitic and sidereal corrections for the S2 star due to the primary's quadrupole as
functions of the $i,~\epsilon$ angles characterizing the spatial orientation of the spin axis of the MBH in the
Galactic center. Lower row: same as in the upper row for the S0-102 star.
}
\label{S2S0102_Q_periods}
\end{figure*}

Similarly to Section~\ref{subsec:LTsec}, also in this case we explore all the quadrupole corrections of
periods as functions of the spatial orientation of the spin axis of the Galactic MBH for the star S2 and S0-102.
The results are showed in Figure~\ref{S2S0102_Q_periods}. We can see that the period
corrections due to the quadrupole have two local minima and two
maxima in the $i-\epsilon$ plane, as it can be inferred from Equation~\ref{eq:q} itself.
For S2, the effects are within the ranges $-2~\textrm{s}\lesssim\delta_Q
P_\textrm{ano}\lesssim1~\textrm{s},~-2.5~\textrm{s}\lesssim\delta_QP_\textrm{dra}
\lesssim2.2~\textrm{s},~ -2.7~\textrm{s}\lesssim\delta_Q P_\textrm{sid}\lesssim 3.3~\textrm{s}$, while for S0-102 we
have$-2.2~\textrm{s}\lesssim\delta_Q P_\textrm{ano}\lesssim 1.1~\textrm{s},~
-15~\textrm{s}\lesssim\delta_QP_\textrm{dra}\lesssim 24~\textrm{s},~ -2.6~\textrm{s}\lesssim\delta_Q
P_\textrm{sid}\lesssim2.3~\textrm{s}$. Thus, the period corrections by quadrupole-induced effects are about orders of
magnitude smaller than those of the Kerr-type, making the detections of them quite challenging.

\section{Discussion}\lb{discussion}
S-stars orbiting the MBH much closer than the currently known ones are expected to be revealed by
future telescopes. Their observables, i.e., the apparent proper motions and the redshifts, can be collected
with high accuracy by such future facilities which can be used to reconstruct the orbital elements of the
target star. Although the projected line of sight distance of the star with respect to the MBH can not be directly
measured, it can be inferred by the accelerations of the proper motions and the redshift: $z=-{\rm
sign}(\ddot{z})\sqrt{MG/|A|-x^2- y^2}$, where $|A|=\sqrt{\ddot x^2+\ddot y^2+\ddot z^2}$, $\ddot
x,~\ddot y,~\ddot z$ are the accelerations in the $x,~y,~z$ directions, and $z$ represents the line of sight direction.
From such measurements, it is straightforward to obtain the corresponding orbital elements and the three periods
according to the measured true anomaly, argument of latitude and true longitude. If the three periods are not
identical to each other, it points towards a strong evidence of the need of the general relativity corrections.

{We notice that these measured periods are combinations of both the Newtonian
and other GR corrections discussed in Section~\ref{risultati}. As we have showed, the period corrections
due to the quadrupole momentum  amount to about $1-10^3$ s even though the
S-star can be as closed as about $10-100~\textrm{AU}$ from the MBH. Considering that
a single exposure time of the S-stars could be comparable to or longer than such time spans,
in the GC environment the quadrupole-induced period corrections  may not be well constrained so to be ignorable. If the orbital elements are measured with considerable accuracy, the sum of the period corrections, given by Equation~\ref{eq:ge},~\ref{eq:lt} or the numerical simulations, can then be used to be compared with the measured three types of periods such that the spin of the MBH is determined. By such an implementation,
the contributions of Kerr effects and the Schwarzschild-type effects are separated, providing independent
constraints on the relativistic effects along with their orbital precession. Furthermore, the
constraints could be improved if the three periods can be measured for more than a single S-star.

However, the orbital elements in Equation~\ref{eq:ge},
~\ref{eq:lt} and~\ref{eq:q} may not be clearly determined as they are defined according to the
local state vectors which can not be measured directly. The practical measurement of them, and
also the related period corrections by the method mentioned above, may
need to be implemented in a more complex framework. For example,
the GR one which incorporates the propagation of light from the star to the observer, and
also includes other complexities that are needed to accurately describe the observed motion of the star
around the MBH. We defer such a realization and comprehensive studies to future works.}

We also notice that the background stars can also induce Newtonian perturbations on the orbital
periods~\citep{2017ApJ...834..198Z}. Their most peculiar signature consists of the fact that the orbital period of the
S-stars changes in every revolution~\citep{2017ApJ...834..198Z}. Thus, it is expected that they should appear
different with respect to those due to the general relativistic effects discussed here, which strongly
suggests that these two effects are separable.

\section{Conclusions}\lb{fine}
In presence of post-Keplerian accelerations, of both Newtonian and post-Newtonian origin, the degeneracy affecting the
draconitic, anomalistic and sidereal orbital periods of a gravitationally bound two-body system is removed. Analytical
calculations of the corresponding post-Keplerian corrections, mainly in the small eccentricity approximation, appeared
recently in the literature. The extensive numerical simulations performed in the present work, applied to various
orbital configurations of a target S--star orbiting the black hole which is supposed to lurk in Sgr
A$^\ast$ at the center of the Galaxy, confirmed the previous results by extending them to highly elliptic orbits.  It
turned out that, for a target star with $a = 300-800$ AU and $e \approx 0.8$, the post-Newtonian Schwarzschild-type
acceleration causes corrections as large as about  $10-100~\textrm{d}$, the Lense-Thirring ones are up to $\approx
10~\textrm{h}$, while the quadrupole-driven ones amount to $100~\textrm{s}$. Furthermore, we considered also the
currently known S2 and S0-102 stars by numerically investigating their periods. In particular, the Schwarzschild-type
anomalistic corrections agreed well with the values provided by the analytical formulas in the literature; furthermore,
the Lense-Thirring and quadrupole-induced orbital time intervals were studied as functions of the spatial orientation
of the spin axis of the hosting black hole.

As possible directions for future work, an extension of the previously published analytical results to large values of
the eccentricity would be desirable in view of a comparison with the present numerical results.
Furthermore, we stress that our results have a general validity, not being restricted just to the Galactic Center; they can
be straightforwardly extended also to other astronomical and astrophysical scenarios, like, e.g., exoplanets, which
could turn out to be promising in view of future potential detections.
Finally, the approach followed here can be extended, in principle, also to various modified models of 
gravity which, in the recent years, have come out to center stage. In view of their huge amount, dealing  with even a fraction of them is outside the scopes of the present paper. Indeed, a comprehensive treatment of their impact on the orbital periods would deserve one or more dedicated papers, both of analytical and numerical nature. Whatever the post-Keplerian effect one is interested in, it is hoped that the present analysis may prompt dedicated studies by astronomers aimed to testing the actual measurability of the orbital periods considered here and determining the associated accuracy.

\acknowledgements
\noindent  We thank the anonymous referee for her/his insightful suggestions which contributed to ameliorate the 
manuscript. This work was supported in part by the National Natural Science Foundation of China under grant 
Nos. 11603083, the Fundamental Research Funds for the Central Universities grand No. 161GPY51.


\end{document}